\documentclass[aps,prl,showpacs,amsmath,amssymb,amsfonts,superscriptaddress,twocolumn,lengthcheck]{revtex4-2}
\usepackage{color}
\usepackage{ulem}
\usepackage{amsmath}
\usepackage{amssymb}
\usepackage{graphics}
\usepackage{graphicx}
\usepackage{dcolumn}
\usepackage[pdfpagemode=UseNone,pdfstartview=FitH,colorlinks=true,linkcolor=blue,urlcolor=blue,anchorcolor=blue,citecolor=blue]{hyperref}
\usepackage{lineno}
\usepackage{booktabs}

\begin{document}
\title{Exploring Dynamics of Open Quantum Systems in Naturally Inaccessible Regimes}

\author{Xu-Ke Gu}
\affiliation{Zhejiang Key Laboratory of Micro-Nano Quantum Chips and Quantum Control, School of Physics, and State Key Laboratory for Extreme Photonics and Instrumentation, Zhejiang University, Hangzhou 310027, China}

\author{Li-Zhou Tan}
\affiliation{Zhejiang Key Laboratory of Micro-Nano Quantum Chips and Quantum Control, School of Physics, and State Key Laboratory for Extreme Photonics and Instrumentation, Zhejiang University, Hangzhou 310027, China}

\author{Franco Nori}
\affiliation{Theoretical Quantum Physics Laboratory, Cluster for Pioneering Research, RIKEN, Wakoshi, Saitama 351-0198, Japan} \affiliation{Center for Quantum Computing, RIKEN, Wakoshi, Saitama, 351-0198, Japan} \affiliation{Quantum Research Institute, The University of Michigan, Ann Arbor, Michigan 48109-1040, USA}

\author{J. Q. You}
\email[Corresponding author.~Email:~]{jqyou@zju.edu.cn}
\affiliation{Zhejiang Key Laboratory of Micro-Nano Quantum Chips and Quantum Control, School of Physics, and State Key Laboratory for Extreme Photonics and Instrumentation, Zhejiang University, Hangzhou 310027, China}
\affiliation{College of Optical Science and Engineering, Zhejiang University, Hangzhou 310027, China}	

\date{\today}

\begin{abstract}
	Markovian open quantum systems are governed by the Lindblad master equation where the dissipation contains two parts, i.e., the anti-Hermitian operator and the quantum jumps, which share a common dissipation rate. We generalize the Lindblad master equation via postselection to a generalized Liouvillian formalism in which the effective damping rate of the anti-Hermitian operator can be different from the quantum jump rate. Our formalism provides a parameter space with regimes inaccessible in naturally-occurring systems. We explore these new regimes and find several interesting results including negative damping rates and generalized Liouvillian exceptional points. In a previously unexplored zero-damping Liouvillian regime where the damping rate is negligible, we investigate the effect only due to the quantum jumps and show an unusual polynomial decay of the excited state. This generalized Liouvillian formalism offers opportunities to explore novel phenomena and quantum technologies associated with the peculiar behavior of quantum jumps.
\end{abstract}


\maketitle
{\it Introduction.}---The dynamics of open quantum systems plays a pivotal role in decoherence theory, quantum measurement, and quantum information. The most well-known description harnesses the master equation $\dot{\rho}=\mathcal{L}\rho$, where $\mathcal{L}$ is a superoperator called {\it Liouvillian}. In the commonly-studied Markovian dynamics, $\mathcal{L}$ is of the Lindblad form~\cite{Lindblad-CMP-76,Scully-Cambridge-97,Agarwal-Cambridge-12,Lidar-arxiv-19}, which contains a Hermitian Hamiltonian $\hat{H}$ for unitary evolution and a so-called Lindblad dissipator $\mathcal{D}$ resulting from the environmental influences. The dissipator can be divided into the anti-Hermitian operator $\hat{A}$ describing continuous amplitude damping and the quantum jump $\mathcal{J}$ presenting abrupt stochastic transitions~\cite{Nagourney-PRL-86,Vijay-PRL-11,Murch-Nature-13,Weber-Nature-14,Minev-Nature-19}. Moreover, $\hat{A}$ can be incorporated into $\hat{H}$ to give a non-Hermitian Hamiltonian (NHH)~\cite{Minganti-PRA-19}. Therefore, the Lindblad Liouvillian (LL) can be interpreted as a combination of both NHH and quantum jumps. While quantum jumps are crucial in the quantum regime~\cite{Minganti-PRA-19,Arkhipov-PRA-20,Minganti-PRA-20,Gneiting-PRA-21,Gneiting-PRR-22}, these may be neglected in the macroscopic circumstances well described by semiclassical approaches~\cite{El-Ganainy-NP-18,Minganti-PRA-19}. In this respect, the NHH can be regarded as the semiclassical approximation of the more general LL. Recently, extensive researches on NHH dynamics in open systems have yielded significant insights into a range of intriguing phenomena, including exceptional points (EPs), non-reciprocal light transmission, and topological phases (see, e.g., \cite{Feng-NPhot-17,El-Ganainy-NP-18,Miri-Science-19,Ozdemir-NM-19,Ashida-AIP-20,Bergholtz-RMP-21}).

\begin{figure}
	\centering
	\includegraphics[width=0.48\textwidth]{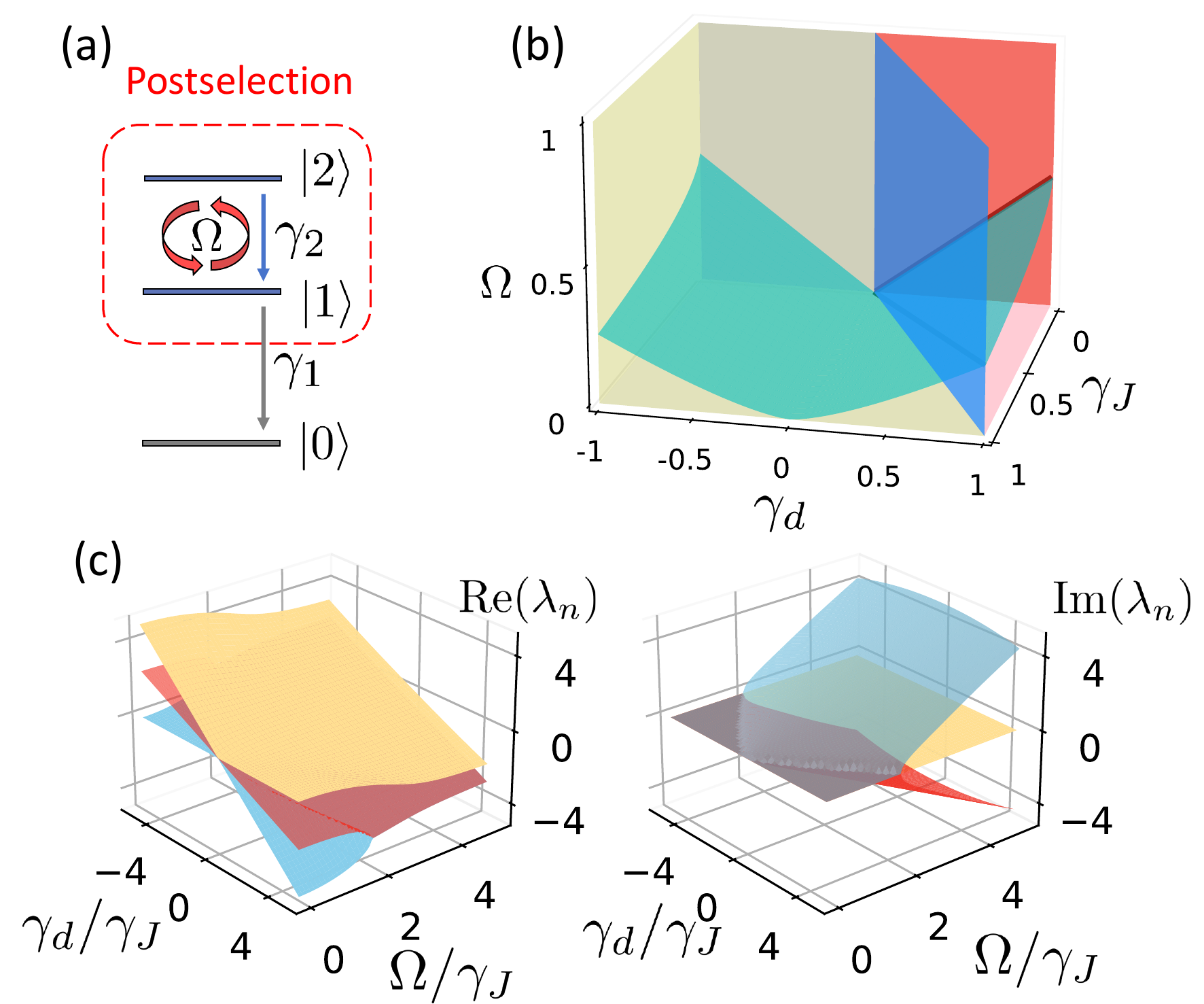}%
	\caption{(a) A ladder-type three-level system to implement a qubit governed by the generalized Liouvillian via postselection. (b) The 3D parameter space related to the generalized Liouvillian in Eq.~(\ref{GL_mat}). (c) Three of the four eigenvalues of the generalized Liouvillian in Eq.~(\ref{GL_mat}): $\lambda_0$ (yellow), $\lambda_1$ (blue), and  $\lambda_2$ (red).}
	\label{fig1}
\end{figure}

The emergence of a NHH from the LL implies the possibility to relax an implicit restriction imposed by the LL: that the {\it damping} rate $\gamma_d$ of $\hat{A}$ equals the {\it quantum jump} rate $\gamma_J$ of $\mathcal{J}$ in naturally-occurring systems. If separate controls of $\gamma_d$ and $\gamma_J$ become realizable via an artificial system, we can expand the accessible regimes from the LL ($\gamma_d=\gamma_J$) and NHH ($\gamma_J=0$), i.e., the blue and red planes in Fig.~\ref{fig1}(b), to a 3D parameter space. Indeed, the physical realization of this extension is rather challenging. A hybrid-Liouvillian formalism~\cite{Minganti-PRA-20} was proposed to effectively tune $\gamma_J$ via postselection. This formalism makes the $\alpha\equiv\gamma_d/\gamma_J\in[1,+\infty)$ region [pink region in Fig.\ref{fig1}(b)] available; but the region $\alpha\in[0,1)$ remains elusive. In this Letter, we develop a new approach to generalize the LL by introducing an effective damping rate. Our generalized Liouvillian formalism can reach the previously unexplored $\alpha\in(-\infty,1)$ region [yellow region in Fig.~\ref{fig1}(b)], including the novel regime with negative damping.

Numerous interesting results have been obtained when exploring only the NHH plane. Thus, additional novel phenomena are expected in the 3D parameter space with regimes which might be better than those in natural systems for quantum information processing. We discover that the negative damping enhances quantum-jump-induced state-mixing, in contrast to its positive counterpart. We reveal the {\it disappearance} and {\it reappearance} of EPs when tuning $\alpha$ across 0. An EP is a branch-point singularity in the parameter space of a system at which two or more eigenvalues and their eigenvectors coalesce~\cite{Miri-Science-19,Ozdemir-NM-19,Guo-PRL-09,Ruter-NP-10,Peng-NP-14,Zhang-NC-15,Zhang-PRXq-21,Arkhipov-PRA2-20}. It has various applications, including lasing~\cite{Brandstetter-NC-14,Peng-Science-14,Peng-PNAS-16,Wong-NPhot-16}, unidirectional invisibility~\cite{Lin-PRL-11,Regensburger-Nature-12,Feng-NM-13,Fleury-NC-15}, and topological mode transfer~\cite{Xu-Nature-16,Doppler-Nature-16,Choi-NC-17,Zhang-PRX-18}. We also investigate a peculiar regime of zero-damping Liouvillian, where the damping rate is negligible and the effect only due to the quantum jumps is explored for the first time~\cite{Minganti-PRA-20}. Indeed, we find an unusual {\it polynomial} decay that is experimentally accessible.

{\it Generalized Liouvillian.}---In the Lindbald master equation $\dot{\rho}=\mathcal{L}\rho$, the Liouvillian $\mathcal{L}$ has the following form~\cite{Lindblad-CMP-76,Scully-Cambridge-97,Agarwal-Cambridge-12,Lidar-arxiv-19}:
\begin{equation}\label{LL}
	\begin{split}
		\mathcal{L}\rho&=-i[\hat{H},\rho]+\sum_k\gamma_k\big[\hat{\Gamma}_k\rho \hat{\Gamma}_k^\dagger-\frac{1}{2}(\hat{\Gamma}_k^\dagger \hat{\Gamma}_k\rho+\rho \hat{\Gamma}_k^\dagger \hat{\Gamma}_k)\big]\\
		&\equiv-i[\hat{H},\rho]-i\{\sum_k\gamma_k\hat{A}_k,\rho\}+\sum_k\gamma_k\mathcal{J}(\hat{\Gamma}_k)\rho,
	\end{split}
\end{equation}
where $\hat{H}$ and $\rho$ are the Hamiltonian and the density matrix of the system, respectively. The Lindblad operators $\hat{\Gamma}_k$ are associated with the dissipations characterized by the anti-Hermitian operator $\hat{A}_{k}\equiv-i\hat{\Gamma}_k^\dagger \hat{\Gamma}_k/2$ and the quantum jump superoperator $\mathcal{J}(\hat{\Gamma}_k)\rho\equiv\hat{\Gamma}_k\rho \hat{\Gamma}_k^\dagger$. Clearly, $\hat{A}_{k}$ and $\mathcal{J}(\hat{\Gamma}_k)$ share a common rate $\gamma_k$, corresponding to the blue plane in Fig.~\ref{fig1}(b). If quantum jumps are neglected, Eq.~(\ref{LL}) is reduced to $\mathcal{L}\rho=-i(\hat{H}_{\rm eff}\rho-\rho\hat{H}^{\dag}_{\rm eff})$, with $\hat{H}_{\rm eff}=\hat{H}+\sum_k\gamma_k\hat{A}_k$ being the NNH, which corresponds to the red plane in Fig.~\ref{fig1}(b).

Now we generalize the LL by introducing an artificial system, i.e., an {\it effective} two-level system reduced from a ladder-type three-level system $\{\vert 0\rangle,\vert 1\rangle,\vert 2\rangle\}$ via postselection [Fig.~\ref{fig1}(a)], where the LL in Eq.~(\ref{LL}) takes two Lindblad operators $\hat{\Gamma}_1=\vert 0\rangle\langle 1\vert$ and $\hat{\Gamma}_2=\vert 1\rangle\langle 2\vert$. The postselection refers to the process of selecting and analyzing only a subset of measurement outcomes that satisfy specific criteria, hence affecting the quantum system retroactively. Here we can isolate the dynamics in the $\{\vert 1\rangle,\vert 2\rangle\}$ subspace by postselecting experiment trials where the system never undergoes a quantum jump $\mathcal{J}(\Gamma_1)$~\cite{Naghiloo-NP-19}. The resulting dynamics confined in that subspace is governed by $\dot{\rho}_{\rm sub}=(-\gamma_1+\mathcal{L}_g)\rho_{\rm sub}$, with (see Sec.~I in \cite{SM})
\begin{equation}\label{GL}
	\mathcal{L}_g\rho_{\rm sub}=-i[\hat{H}_{\rm sub},\rho_{\rm sub}]-i\{\gamma_d\hat{A}_2,\rho_{\rm sub}\}+\gamma_J\mathcal{J}(\hat{\Gamma}_2)\rho_{\rm sub},
\end{equation}
where $\hat{H}_{\rm sub}$ and $\rho_{\rm sub}$ are the reduced Hamiltonian and density matrix in that subspace, the quantum jump rate is $\gamma_J\equiv\gamma_2$, and the effective damping rate $\gamma_d\equiv\gamma_2-\gamma_1$ is defined as the damping rate of $\vert 2\rangle$ relative to $\vert 1\rangle$.
Thus,
\begin{equation}\label{sub_rhot}
	\rho_{\rm sub}(t)=\exp{(-\gamma_1 t)}\exp{(\mathcal{L}_g t)}~\rho_{\rm sub}(0).
\end{equation}
As shown in Eq.~(\ref{sub_rhot}), we extract $\gamma_1$ to be a global decay, characterizing the probability leakage from the subspace to the outside. This is shown by exponentially decreasing the number of selected experiment trials in Sec.~IV of~\cite{SM}. In fact, it is the damping-rate difference between $\vert 2\rangle$ and $\vert 1\rangle$ (i.e., the effective damping rate $\gamma_d$) that competes with the coupling and the quantum jump to determine the dynamics inside the subspace. In NHH research, such a role of the relative damping has already been utilized to realize passive PT-symmetry systems without any gain~\cite{Feng-NPhot-17,Ozdemir-NM-19,Guo-PRL-09}.

Due to the postselection, $\rho_{\rm sub}$ is not trace-preserving, i.e., ${\rm Tr}[\rho_{\rm sub}(t)]\ne 1$, so it needs to normalize the state~\cite{Gisin-JPAMG-81,Graefe-JPAMT-10,Brody-PRL-12}:
\begin{equation}\label{norm}
	\tilde{\rho}_{\rm sub}(t)=\frac{\rho_{\rm sub}(t)}{{\rm Tr}[\rho_{\rm sub}(t)]}=\frac{e^{\mathcal{L}_g t}\rho_{\rm sub}(0)}{{\rm Tr}[e^{\mathcal{L}_g t}\rho_{\rm sub}(0)]}.
\end{equation}
The normalized density matrix $\tilde{\rho}_{\rm sub}(t)$ should be regarded as the physical state of the system, in the sense that it provides the correct measurement statistics~\cite{Nielsen-Cambridge-12}. In practice, to measure a single quantum system, multiple identical trials are conducted, forming an ensemble of the quantum system. Quantum trajectory theory~\cite{Carmichael-Springer-93,Dalibard-PRL-92,Gardiner-PRA-92,Plenio-RMP-98} describes the evolution of the system in each trial as a stochastic quantum trajectory, and measurement results are obtained by averaging over all trajectories. However, postselection only selects specific trials to average, normalizing their probabilities. Thus, measurement results for the $\{\vert 1\rangle,\vert 2\rangle\}$ subspace after postselection are correctly predicted by the normalized state $\tilde{\rho}_{\rm sub}(t)$.

Moreover, it follows from Eq.~(\ref{norm}) that the evolution
\begin{equation}\label{sub_evo}
	\rho'_{\rm sub}(t)= \exp{(\mathcal{L}_g t)}~\rho'_{\rm sub}(0),
\end{equation}
with $\rho'_{\rm sub}(0)=\rho_{\rm sub}(0)$, gives the same normalized state $\tilde{\rho}_{\rm sub}(t)$ as $\rho_{\rm sub}(t)$. Thus, we can eliminate the decay factor $\exp{(-\gamma_1t)}$ and only consider Eq.~(\ref{sub_evo}) to describe the dynamics confined in the subspace. Hence, the generator of the dynamics in the $\{\vert 1\rangle,\vert 2\rangle\}$ subspace is our generalized Liouvillian $\mathcal{L}_g$. This is equivalent to constructing a qubit $\{\vert 1\rangle,\vert 2\rangle\}$ with the damping rate $\gamma_d$ different from the quantum jump rate $\gamma_J$ [see Fig.\ref{fig1}(a)]. With such a qubit, we can explore the naturally inaccessible regimes $\alpha\in(-\infty,1)$ [the yellow region in Fig.\ref{fig1}(b)] by controlling the ratio between $\gamma_1$ and $\gamma_2$. Such control can be achieved following the method in Ref.~\cite{Naghiloo-NP-19}, which is based on the Purcell effect~\cite{Purcell-PR-46}.

For a further look into $\mathcal{L}_g$, we need to specify the Hamiltonian $\hat{H}_{\rm sub}$. Here we consider a coherently driven qubit in the rotating frame: $\hat{H}_{\rm sub}=\frac{\Omega}{2}(\vert 2\rangle\langle 1\vert+\vert 1\rangle\langle 2\vert)$, where $\Omega$ is the drive amplitude [see Fig.~\ref{fig1}(a)]. The superoperator $\mathcal{L}_g$ can be expressed in a matrix form as (see Sec.~II in \cite{SM})
\begin{equation}\label{GL_mat}
	\mathcal{L}_g=\left(\begin{matrix}
		0 & i\frac{\Omega}{2} & -i\frac{\Omega}{2} & \gamma_J\\
		i\frac{\Omega}{2} & -\frac{\gamma_d}{2} & 0 & -i\frac{\Omega}{2}\\
		-i\frac{\Omega}{2} & 0 & -\frac{\gamma_d}{2} & i\frac{\Omega}{2}\\
		0 & -i\frac{\Omega}{2} & i\frac{\Omega}{2} & -\gamma_d
	\end{matrix}\right).
\end{equation}
The eigenvalues of $\mathcal{L}_g$ are $\{\lambda_n,n=0,1,2\}$ as shown in Fig.~\ref{fig1}(c) (see Sec.~II in \cite{SM} for explicit expressions) and $\lambda_3=-\frac{\gamma_d}{2}$. The generically non-Hermitian $\mathcal{L}_g$ has both right eigenmatrices $\rho_i$, defined by $\mathcal{L}_g\rho_i=\lambda_i\rho_i$, and left eigenmatrices $\sigma_i$, defined by $\mathcal{L}_g^\dagger\sigma_i=\lambda_i^*\sigma_i$. The left and right eigenmatrices are {\it biorthonormal} in the sense that one can always normalize them to have inner products ${\rm Tr}(\sigma_i^\dagger\rho_j)=\delta_{ij}$. Thus, if $\mathcal{L}_g$ is diagonalizable, the evolution $\rho_{\rm sub}(t)$ can be solved in terms of eigenmatrices as
\begin{equation}\label{evo}
		\rho_{\rm sub}(t)=\sum_{i=0}^{3} \exp{(\lambda_it)}{\rm Tr}[\sigma_i^\dagger\rho_{\rm sub}(0)]\rho_i.
\end{equation}

\begin{figure}
	\includegraphics[width=0.48	\textwidth]{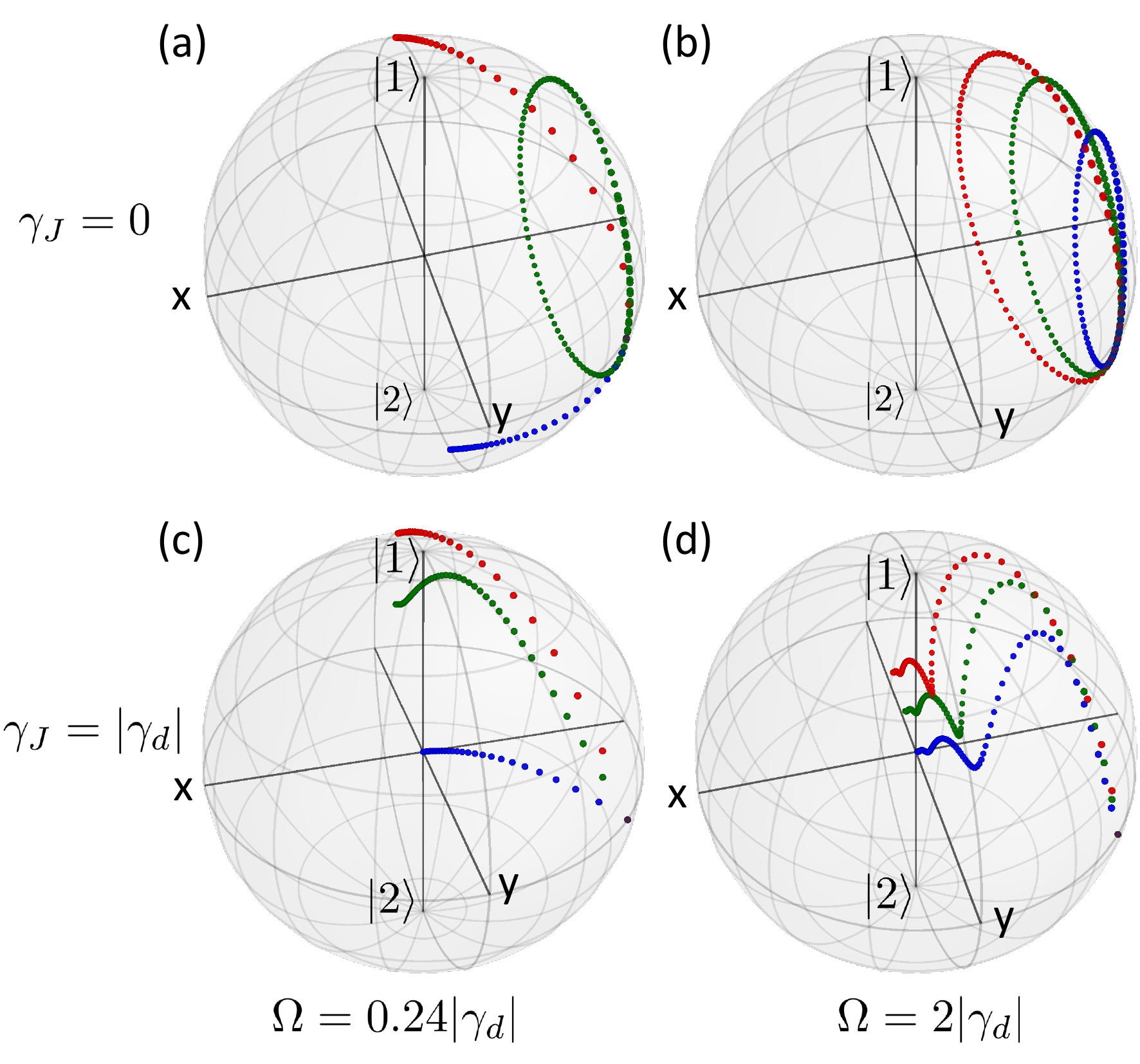}%
	\caption{Evolution of the initial state $\frac{1}{\sqrt{2}}(\vert 1\rangle+e^{i\frac{3\pi}{4}}\vert 2\rangle)$ under negative damping (blue) compared with those under positive damping (red) and zero damping (green). (a) $\gamma_J=0$, $\Omega=0.24|\gamma_d|$; (b) $\gamma_J=0$, $\Omega=2|\gamma_d|$; (c) $\gamma_J=|\gamma_d|$, $\Omega=0.24|\gamma_d|$; (d) $\gamma_J=|\gamma_d|$, $\Omega=2|\gamma_d|$. The evolution results are obtained using Eq.~(\ref{evo}) and visualized in Bloch spheres using QuTiP~\cite{Johansson-CPC-2012,Johansson-CPC-2013}.}
	\label{fig2}
\end{figure}

{\it Negative damping rates.}---The effective damping rate is defined as the damping rate of the upper state relative to the lower state, $\gamma_d\equiv\gamma_2-\gamma_1$, which becomes negative when $\gamma_1>\gamma_2$. The negative (positive) damping rate has a clear physical~interpretation: that the probability amplitude of the upper state $\vert 2\rangle$ increases (decreases) with time. This leads to an evolution favoring $\vert 2\rangle$ ($\vert 1\rangle$) shown as the blue (red) dots in Fig.~\ref{fig2}(a) where $\gamma_J=0$ and $\Omega=0.24|\gamma_d|$. In fact,~a similar evolution to a steady state near the upper state of a qubit has been experimentally demonstrated in Ref.~\cite{Naghiloo-NP-19}, confirming the negative damping.

Rather than simply differing in evolution direction when quantum jumps are neglected, negative damping exhibits a fundamental distinction from positive damping when in competition with the quantum jumps. As shown in Fig.~\ref{fig2}(c), where $\gamma_J=|\gamma_d|$ and $\Omega=0.24|\gamma_d|$, the qubit under positive damping evolves to a steady state near $\vert 1\rangle$ and is only slightly mixed during the evolution, whereas the qubit under negative damping evolves to the fully mixed state. The negative damping enhances the quantum-state mixing induced by quantum jumps. Intuitively, the negative damping increases the probability of $\vert 2\rangle$, so more quantum jumps $\mathcal{J}(\Gamma_2)$ happen, i.e., more state-mixing.

Moreover, the situation is different at the other side of the EP, as shown in Fig.~\ref{fig2}(b) and (d), where $\Omega=2|\gamma_d|$. The effect of the damping is to shift the no-damping path (green), and the sign of the damping rate determines the shifting direction. In Fig.~\ref{fig2}(d), for~the naturally accessible LL regime ($\gamma_d=\gamma_J$), the qubit spirally evolves to the steady state $(0,-2\gamma_d\Omega,\gamma_d^2)/(\gamma_d^2+2\Omega^2)$ in the Bloch sphere (see Sec.~II in \cite{SM}), surprisingly left with some information of the dissipation $\gamma_d$ and the coupling $\Omega$. Whereas for the negative LL ($\gamma_d=-\gamma_J$), the qubit evolves to a fully mixed state, losing all the information.

{\it Exceptional points and zero-damping Liouvillian.}---With the qubit governed by $\mathcal{L}_g$, we can now investigate the generalized Liouvillian EPs. For $\mathcal{L}_g$ in Eq.~(\ref{GL_mat}), two eigenvalues $\lambda_1$ and $\lambda_2$, as well as their eigenmatrices, coalesce at the merging line in Fig.~\ref{fig1}(c). The explicit condition for the generalized Liouvillian EPs can be derived as (see Sec.~III in \cite{SM})
\begin{equation}\label{GLEP}
	108\gamma_J^2\Omega^4+\left(4\Omega^2-\gamma_d^2\right)^3=0,\ ~~~\Omega\ge 0	
\end{equation}
This generalized condition for EPs is shown as the cyan surface in Fig.~\ref{fig1}(b), which is symmetric about $\alpha=0$ because Eq.~(\ref{GLEP}) is invariant under the sign reversal of $\gamma_d$. When $\gamma_J=0$ ($\gamma_J=\gamma_d$), Eq.~(\ref{GLEP}) is reduced to the special case $\Omega=\gamma_d/2$ ($\Omega=\gamma_d/4$), which is exactly the condition for the EPs of the NHH (LL) shown as the black intersecting line of the generalized Liouvillian EP surface and the NHH (LL) plane in Fig.~\ref{fig1}(b).

The EPs of the NHH can emerge from the Liouvillian EPs when changing $\alpha$ from 1 to $+\infty$, as theoretically studied using the hybrid-Liouvillian formalism~\cite{Minganti-PRA-20}. However, in the generalized Liouvillian formalism based on the ladder-type three-level system, we can tune $\alpha$ in the region $(-\infty,1)$ to have it across 0 to observe the disappearance and reappearance of EPs. We say EPs disappear at $\alpha=0$ in the sense that Eq.~(\ref{GLEP}) reduces to $\Omega=0$ and the EP-associated phase transition can no longer be observed. This peculiar $\alpha=0$ regime, where the damping rate is negligible , also enables us to study the dissipative effect only due to quantum jumps, which is a pure quantum effect associated with the zero-damping Liouvillian.

\begin{figure}
	\includegraphics[width=0.48	\textwidth]{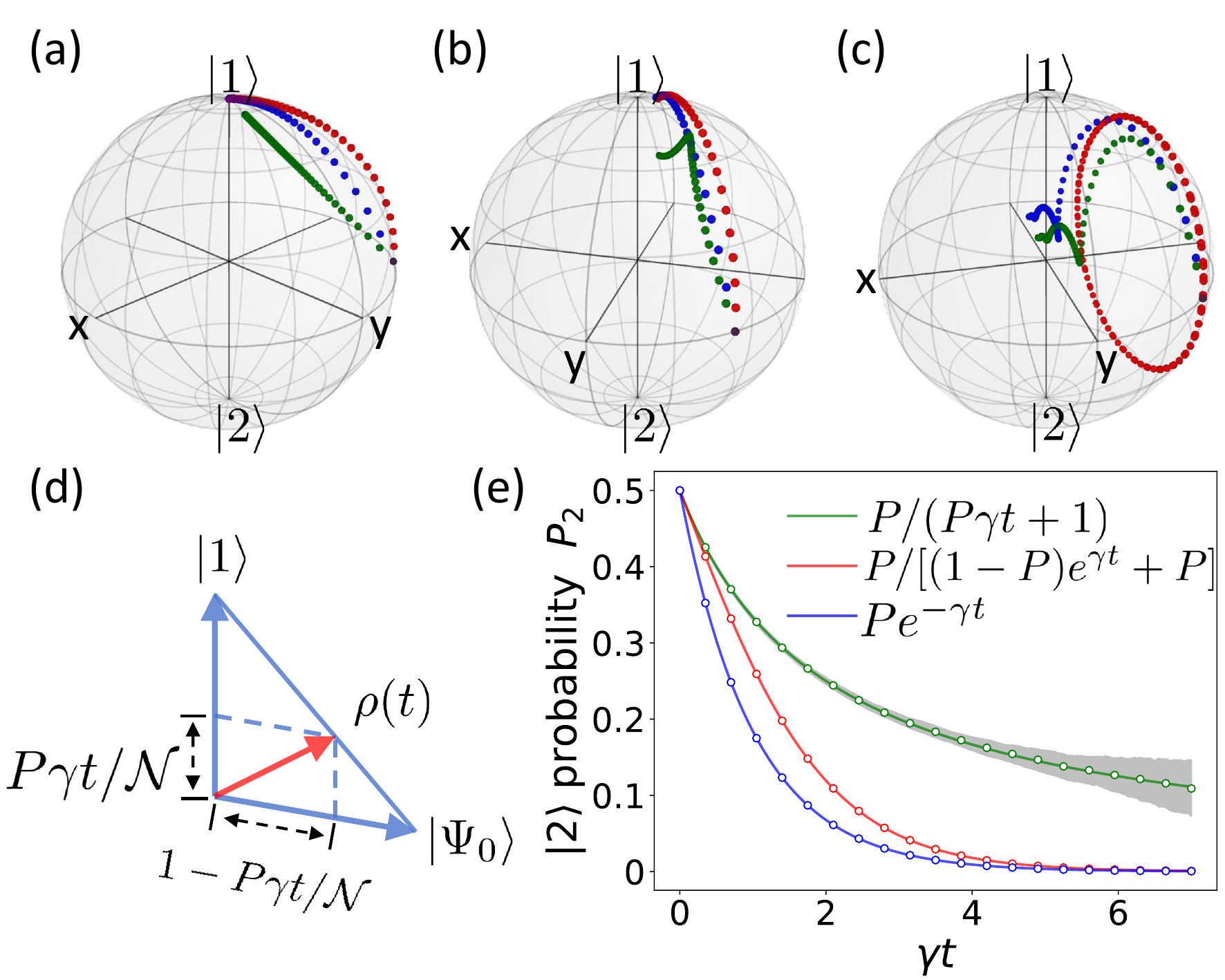}%
	\caption{(a)--(c) Dynamics in the zero-damping Liouvillian (ZDL) regime (green) compared with the dynamics of the NHH (red) and LL (blue) under $\Omega=0,0.24\gamma$ and $2\gamma$, respectively. The evolution results are obtained using Eq.~(\ref{evo}) and visualized in Bloch spheres. (d) Quantum jumps mix $\vert\Psi_0\rangle$ with $\vert 1\rangle$, resulting in a straight line evolution. (e) Probability decay of $\vert 2\rangle$ corresponding to the~evolution in (a). The circles for the LL (blue) and NHH (red) are obtained by standard master equation simulations. The circles for the ZDL (green) are obtained by Monte-Carlo simulations with $10^4$ quantum trajectories and postselection. The shaded area represents the error. Here simulations are performed using QuTiP~\cite{Johansson-CPC-2012,Johansson-CPC-2013}.}
	\label{fig3}
\end{figure}

Using the initial state $\vert\Psi_0\rangle=\frac{1}{\sqrt{2}}(\vert 1\rangle+e^{i\frac{3\pi}{4}}\vert 2\rangle)$, we now study the zero-damping Liouvillian dynamics ($\gamma_J=\gamma$, $\gamma_d=0$) compared with the dynamics of the NHH ($\gamma_J=0$, $\gamma_d=\gamma$) and the LL ($\gamma_J=\gamma_d=\gamma$) under $\Omega=0,0.24\gamma$ and $2\gamma$, as shown in Fig.~\ref{fig3}(a)-(c), respectively.

In particular, sole dissipative effects at $\Omega=0$ are demonstrated in Fig.~\ref{fig3}(a). For the NHH, the state stays pure when evolving to $\vert 1\rangle$ along an arc on the surface of the Bloch sphere. Whereas for the zero-damping Liouvillian, the state evolves to $\vert 1\rangle$ along a straight line inside the Bloch sphere, indicating the process where the quantum jump $\mathcal{J}(\Gamma_2)$ induces $\vert\Psi_0\rangle$ to mix with $\vert 1\rangle$. The resulting mixed state is
\begin{equation}\label{rhot_ZDL}
	\rho(t)=\frac{1}{\mathcal{N}}(P\gamma t\vert 1\rangle\langle 1\vert+\vert\Psi_0\rangle\langle\Psi_0\vert),
\end{equation}
where $P\equiv\big|\langle 2\vert\Psi_0\rangle\big|^2$ is the probability of the initial state at $\vert 2\rangle$ and $\mathcal{N}=1+P\gamma t$ is the normalization factor. Such mixing is equivalent to a vector addition in the Bloch sphere representation, as shown in Fig.~\ref{fig3}(d); and $\rho(t)$ indeed falls on the straight line defined by $\vert\Psi_0\rangle$ and $\vert 1\rangle$. For the LL, the state evolves along a curve between the NHH arc and the zero-damping Liouvillian line, which can be understood as the interplay between the anti-Hermitian operator and the quantum jumps. The straight line evolution induced by the quantum jumps can be experimentally confirmed using quantum state tomography.

Interestingly, the probability decay of $\vert 2\rangle$, which is much easier to measure, can also demonstrate certain peculiar aspects of the quantum jump effect. As shown in Fig.~\ref{fig3}(e), the zero-damping Liouvillian exhibits a {\it polynomial} decay $P_2(t)=(\gamma t+P^{-1})^{-1}$, which is unusual in contrast to the common exponential decay in both LL and NHH cases. This {\it polynomial} decay can be directly obtained from Eq.~(\ref{rhot_ZDL}), which contains state mixing and normalization. Physically, one can interpret the polynomial decay as a result of both quantum jumps and postselection. In particular, the results of the zero-damping Liouvillian in Fig.~\ref{fig3}(e) are obtained by averaging over quantum trajectories postselected from $10^4$ simulation trials (see Sec.~IV in \cite{SM}). Such a sub-ensemble of the simulated quantum trajectories highly resembles the sub-ensemble formed by postselecting single experimental trials in practice~\cite{Naghiloo-NP-19}. Thus, the polynomial decay is reliable and experimentally accessible.

{\it Discussion and conclusions.}---In the study above, we propose to realize generalized Liouvillians with the~previously unexplored region of $\alpha\in(-\infty,1)$, using a qubit reduced from a ladder-type three-level system via postselection, where the negative and zero damping regimes are inaccessible in natural systems and exhibit novel quantum dynamics. On the other hand, using a qubit reduced from a $\Lambda$-type three-level system via postselection (see Sec.~I in \cite{SM}), we can realize a generalized Liouvillian with $\alpha\in[1,+\infty)$, which can also be realized using the hybrid-Liouvillian formalism~\cite{Minganti-PRA-20}. Thus, we can achieve the generalized Liouvillian of a qubit in the whole parameter region $\alpha\in(-\infty,+\infty)$.

In summary, we have developed a generalized Liouvillian formalism to relax the restriction on dissipation rates imposed by the Lindblad master equation. The effective damping rate, different from the quantum jump rate, leads to an additional dimension in the parameter space. We discover the enhancement of quantum-state mixing by negative damping. We derive the symmetric generalized Liouvillian EP that gradually disappears when $\alpha$ approaches 0. We explore the zero-damping Liouvillian regime to investigate the effect only due to the quantum jumps~and find out the straight line evolution in the Bloch sphere corresponding to a polynomial decay of the upper state. The generalized Liouvillian formalism opens up a vast parameter space, especially regimes where quantum jumps dominate. Quantum jumps are a peculiar quantum behavior without classical counterpart. It can be employed to develop quantum technologies associated with quantum feedback control~\cite{Wiseman-Cambridge-10,Zhang-PR-17}. Moreover, the generalized Liouvillian EPs, which contain the impact of quantum jumps, lay the foundation to extend the rich applications of the semiclassical EPs of the NHH~\cite{Miri-Science-19, Ozdemir-NM-19,Wiersig-PR-20,Bergholtz-RMP-21} to pure quantum cases.

\begin{acknowledgments}
We thank Clemens Gneiting for useful comments. This work is supported by the National Key Research and Development Program of China (Grant No.~2022YFA1405200), and the National Natural Science Foundation of China (Grant Nos.~92265202 and 11934010). F.N. is supported in part by the Japan Science and Technology Agency (JST) [via the CREST Quantum Frontiers program Grant No.~JPMJCR24I2, the Quantum Leap Flagship Program (Q-LEAP), and the Moonshot R\&D Grant No.~JPMJMS2061], and the Office of Naval Research (ONR) Global (Grant No. N62909-23-1-2074).
\end{acknowledgments}

\end{document}